\documentclass[1p,number]{elsarticle}

\usepackage{graphicx}
\usepackage{subfig}
\usepackage{amsmath}
\usepackage{amssymb}
\usepackage[utf8]{inputenc}
\usepackage{url}
\usepackage[bookmarks=false]{hyperref}
\usepackage{graphicx}

\usepackage{program}
\usepackage[table]{xcolor}
\usepackage{multirow}
\usepackage{fixltx2e}

\definecolor{tabbg1}{rgb}{0.85,0.85,0.85}
\definecolor{tabfg1}{rgb}{0.0,0.0,0.0}
\definecolor{tabbg2}{rgb}{0.95,0.95,0.95}
\definecolor{tabfg2}{rgb}{0.0,0.0,0.0}

\usepackage{makecell}
\newlength\hrulethickness
\newcommand{\hlinethick}{\Xhline{\hrulethickness}}

\usepackage[]{color}

\makeatletter
\def\ps@pprintTitle{%
 \let\@oddhead\@empty
 \let\@evenhead\@empty
 \def\@oddfoot{}%
 \let\@evenfoot\@oddfoot}
\makeatother

\begin{document}

\begin{abstract}
The growing gap between processor and memory speeds has lead to complex memory hierarchies as processors evolve to mitigate such divergence by exploiting the locality of reference.
In this direction, the BSC performance analysis tools have been
recently extended to provide insight into the application memory accesses by depicting their temporal and spatial characteristics, correlating with the source-code and the achieved performance simultaneously.
These extensions rely on the Precise Event-Based Sampling (PEBS) mechanism available in recent Intel processors to capture information regarding the application memory accesses.
The sampled information is later combined with the Folding technique to represent a detailed temporal evolution of the memory accesses and in conjunction with the achieved performance and the source-code counterpart.
The reports generated by the latter tool help not only application developers but also processor architects to understand better how the application behaves and how the system performs.
In this paper, we describe a tighter integration of the sampling mechanism into the monitoring package.
We also demonstrate the value of the complete workflow by exploring already optimized state--of--the--art benchmarks, providing detailed insight of their memory access behavior.
We have taken advantage of this insight to apply small modifications that improve the applications' performance.
\end{abstract}

\begin{elskeyword} performance analysis \sep
 memory references \sep
 sampling \sep
 instrumentation
\end{elskeyword}

\begin{frontmatter}

\title{Understanding Memory Access Patterns\\Using the BSC Performance
  Tools}
\tnotetext[]{DOI: 10.1016/j.parco.2018.06.007. \textcopyright 2018
  Elsevier.  This manuscript version is made available under the
  CC-BY-NC-ND 4.0 license \url{http://creativecommons.org/licenses/by-nc-nd/4.0}}
\author[intel]{Harald Servat\corref{cor1}}\ead{harald.servat@intel.com}
\author[bsc,upc]{Jes\'us Labarta}
\author[intel]{Hans-Christian Hoppe}
\author[bsc,upc]{Judit Gim\'enez}
\author[bsc]{Antonio J. Pe\~na}

\cortext[cor]{Corresponding author.}
\address[intel]{Intel Corporation}
\address[bsc]{Barcelona Supercomputing Center (BSC)}
\address[upc]{Universitat Polit\`ecnica de Catalunya (UPC)}

\end{frontmatter}

\section{Introduction}

The growing gap between processor and memory speeds
leads to more and more complex memory hierarchies as
processors evolve generation after generation. The memory
hierarchy is organized in different \textit{strata} to exploit the
applications' temporal and spatial localities of reference. On
one end of the hierarchy lie extremely fast, tiny and power-hungry
registers while on the other end there is the slow,
huge and less energy-consuming DRAM. In between these
two extremes, there are multiple cache levels that mitigate the
expense of bringing data from the DRAM when the application exposes
either spatial or temporal locality. Still, researchers and
manufacturers look for alternatives to improve the memory
hierarchy performance- and energy-wise. For instance,
they consider additional integration directions so that
the memory hierarchy adds layers as scratchpad memories,
stacked 3D DRAM~\cite{3DRAM_Loh} and even non-volatile RAM~\cite{NVMalloc}.

A proper analysis of the application memory
references and its data structures is vital to identify which
application variables are referenced the most, their access cost,
as well as to detect memory streams. All this information might
provide hints to improve the execution behavior by helping
prefetch mechanisms, suggesting on the usage of non-temporal instructions, 
calculating reuse distances, tuning cache organization
and even facilitating research on multi-tiered memory systems.
Two approaches are typically used to address these studies.
First, instruction-based instrumentation tools monitor load/store instructions
and decode them to capture the referenced addresses and
the time to solve the reference. While this approach can capture
all data references and accurately correlates code statements with data references,
it estimates cache access costs by simulating the cache hierarchy
and introduces significant overheads that alter
the observed performance and challenges the analysis with large data
collections and time-consuming analysis, and is thus not
practical for production runs. 
Second, some processors have enhanced their Performance Monitoring Unit
(PMU) to sample memory instructions 
and capture data such as: referenced address, time to solve the reference and
the memory hierarchy level that provides the data.
The sampling mechanisms help to reduce the amount of data captured
and the overhead imposed and thus allow targeting
production application runs. However, the results obtained
using statistical approximations may
require sufficiently long runs to approximate
the actual distribution; still, highly dynamic access patterns or rare performance
excursions may be missed.

The Extrae instrumentation package~\cite{EXTRAE} and the Folding tool~\cite{ServatLGHL11_icpp}
belong to the BSC performance tools suite and have been recently extended
to explore the performance behavior
and the references of the application data objects simultaneously~\cite{ServatLGGL15}.
However, the initial research prototype combined the results of two independent 
monitoring tools (Extrae and the \texttt{perf} tool~\cite{perf}) that monitored the same process
before depicting the results through the Folding tool.
The changes described in this paper address several of the limitations of that prototype.

In this document we describe a fully integrated solution of the initial prototype.
The novelties of this integration include:
\begin{itemize}
\item Simplified the collection mechanism by using the \texttt{perf} kernel infrastructure directly from Extrae to use the Intel Precise Event-Based Sampling (PEBS)~\cite{INTEL_PEBS} mechanism. This avoids to load a kernel module to correlate clocks between the two tools and reduces the overall overhead suffered by the application.
\item Use Extrae capabilities to multiplex load and store instructions in a single application execution. This naturally provides load and store references in a single report while in the prototype it was uneasy due to kernel security features.
\item Extend the Extrae API to create synthetic events that delimit a memory region. This reduces the space needed for intermediate files on applications that allocate data in small consecutive chunks.
\end{itemize}

The organization of this paper is as follows.
Section~\ref{sec:ExtensionsBSCtools} describes the extensions done to the BSC performance tools in order to collect and represent data related to memory data-objects and references to them.
Section~\ref{sec:ApplicationEvaluation} follows with exhaustive performance and memory access analyses of several benchmarks including code modifications and comparing the execution behavior before and after the code changes.
Then Section~\ref{sec:RelatedWork} contextualizes this tool with respect to the state-of-the-art tools.
Finally, Section~\ref{sec:Conclusions} draws conclusions.

\section{Extensions to the BSC performance tools}
\label{sec:ExtensionsBSCtools}

\begin{figure}
	\centering
	\includegraphics[width=0.65\columnwidth]{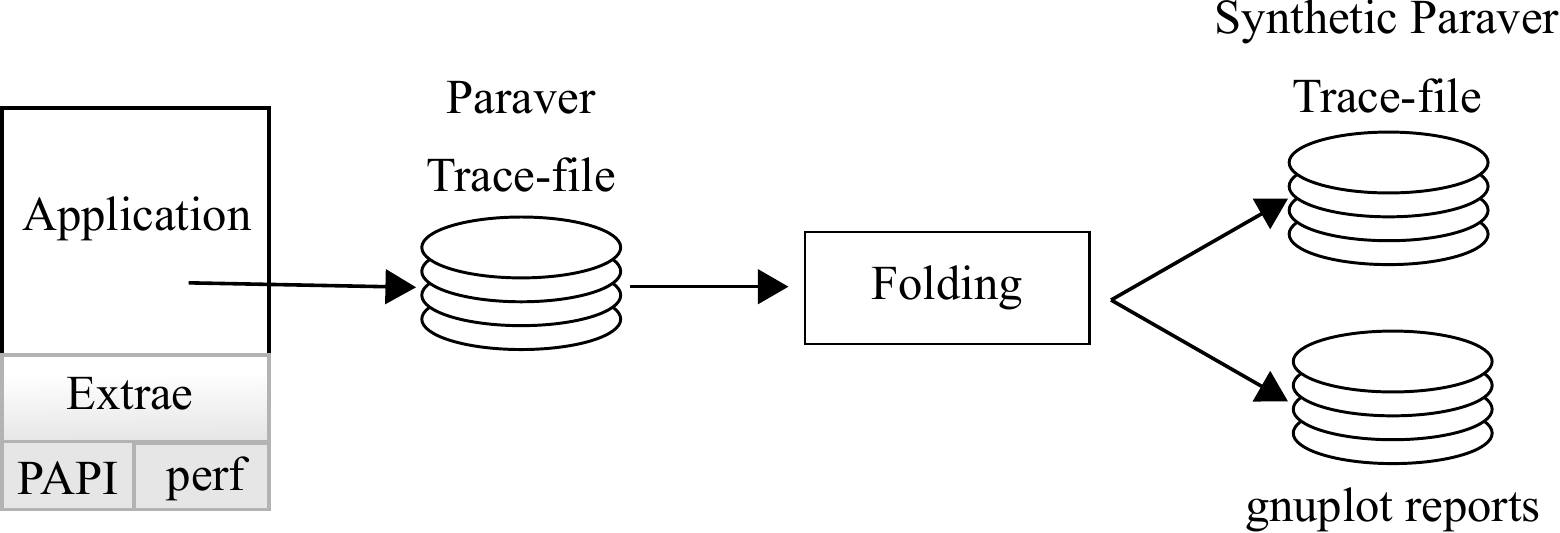}
	\caption{Tool integration for the memory reference analysis.}
	\label{fig:ToolIntegration}
\end{figure}

This section covers the extensions applied to the Extrae and Folding tools.
Figure~\ref{fig:ToolIntegration} depicts the interaction of these tools when exploring a target application.
First, Extrae monitors the target application.
Extrae is an open-source instrumentation and sampling software
which generates Paraver~\cite{Paraver} timestamped event traces for offline analysis.
The package monitors several programming
models (e.g. MPI, OpenMP, OmpSs and POSIX threads)
to allow the analyst to understand the application behavior.
Although Extrae offers an API for manual instrumentation,
it also monitors in-production optimized binaries
through the shared-library preloading mechanisms.
Extrae can also multiplex the performance counters
capturing more performance counters over the application run than
the underlying hardware can collect simultaneously.
The sampling mechanism is implemented on top of time-based alarms 
as well as on top of hardware counters.

After the trace-file has been generated, the Folding tool is invoked.
The Folding tool takes advantage of the repetitive nature of many applications (especially in the HPC environment) and combines sampling and instrumented information to provide detailed progression within a repetitive computing region.
This allows monitoring the application at a coarse sampling frequency without impacting on the application performance.
The Folding generates two different outputs on the delimited repetitive regions.
On the one hand, it generates summarized performance reports that can be explored using the \texttt{gnuplot} tool\footnote{http://gnuplot.info}.
On the other hand, it generates synthetic Paraver trace-files that include all the information from the summarized performance reports and additional details that cannot be represented in the plots.

\subsection{Extensions to Extrae}

\subsubsection{On the collection of the application data-objects}

The modifications on Extrae focus on capturing information of the application data
structures and collecting information about the references to these structures.
Therefore, to help the analyst understand the access patterns,
it is necessary to map addresses to actual application data structures.
Consequently, Extrae has been extended to capture some properties of
static and dynamic variables. With respect to the static
variables, the instrumentation package scans the symbols
within the application binary image using the binutils library\footnote{\url{http://www.gnu.org/software/binutils}}
to acquire their name, starting address and size. Regarding the
dynamically allocated variables, the monitoring package has
been extended to instrument the \texttt{malloc}-related routines.
Extrae captures their input parameters and output results to
determine the starting address and size, as well as a portion
of the call-stack in order to locate them within the user code.
Extrae identifies
them by the top of the call-stack on the allocation site, instead.
Extrae also captures the references to the local (stack)
variables, but the tool cannot track their creation and thus
these references remain unnamed.

As applications may allocate and de-allocate many variables during the
application lifetime, Extrae ignores allocations smaller than a
given threshold (that defaults to 1 MByte but can be
changed by the user) to avoid generating huge trace-files.
This approach limits the analysis when targeting graph- or tree-based or other irregular
applications where allocations may be tiny.
To circumvent this limitation, we have extended the Extrae API to 
create a synthetic events to delimit a memory region based upon begin and end 
addresses.
This approach lets a user wrap small and consecutive dynamic allocations through this API
and correlate the memory references to a synthetic object that represents all the allocations.
As this approach requires manual intervention, one alternative (not currently implemented)
would be to limit the instrumentation a given number of small allocations.

\subsubsection{On the sampling of memory references}

For monitoring the application's memory references,
Extrae uses the PEBS infrastructure.
Despite Extrae relies on PAPI~\cite{PAPI} to collect the value of hardware performance
counters from the PMU, this performance library does capture the PEBS
generated information\footnote{As of the latest released PAPI version (5.5.1)}. 
Consequently, we have modified Extrae to use the \texttt{perf}\footnote{See perf\_event\_open(2) on the Linux manual page.}
subsystem of the Linux kernel to monitor the memory references.
In brief, to configure the PEBS to sample memory references \texttt{perf} has to:
\begin{itemize}
\item allocate a buffer to hold the PEBS samples,
\item setup a \texttt{pe\_event} for a performance counter that captures memory references (e.g. memory operations) and specify a sampling period, and
\item associate an interrupt handler for the interrupts generated when the PEBS buffer is full and processes the PEBS buffer.
\end{itemize}
The reader may wonder on the portability of the monitoring tool to other processors.
We believe that the approach of using the \texttt{perf} subsystem holds true for 
mechanisms similar to PEBS (e.g. IBS in AMD Opteron~\cite{AMD_IBS} and MRK in IBM Power7~\cite{srinivas2011ibm}).
However, we cannot provide specific details.

It is worth to mention that the metrics associated to the memory references depend on the monitored performance counter and within processor families.
For instance, Intel\textsuperscript{\textregistered}~Xeon\textsuperscript{\textregistered}
processors extend PEBS with Load-Latency features that allow monitoring
load instructions and provide the address referenced, the access cost and which part of the memory hierarchy provided the data.
However, store instructions just provide information regarding the address referenced and whether the access hit in cache.

We want to highlight that PEBS records do not contain a time-stamp\footnote{Intel Skylake generation introduces time-stamps in the PEBS records.}.
Also, as we stated before, the PEBS buffer is forwarded to the performance tool when the buffer is full through an interrupt handler.
Since Extrae needs to associate each referenced address a time-stamp, we have taken the approach on Extrae allocating a \mbox{1-entry}
buffer.
When PEBS interrupts the tool after generating each sample, the interrupt handler will associate a time-stamp to it.

\begin{figure}
	\centering
	\includegraphics{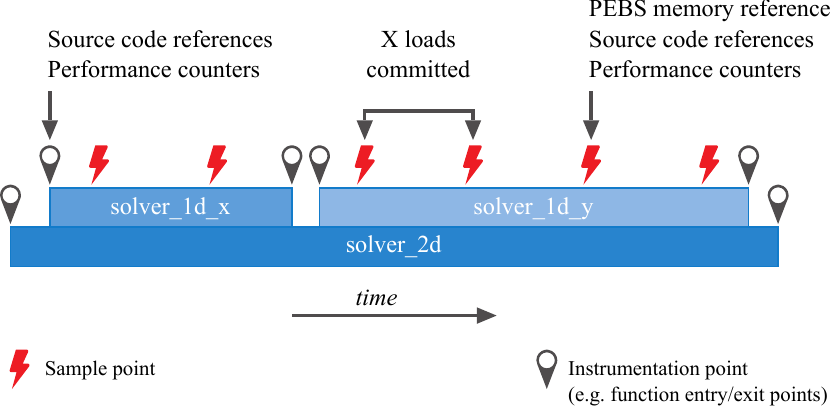}
	\caption{The extensions to the Extrae instrumentation package allow monitoring at instrumentation points as well as PEBS-based points.}
	\label{fig:ExtraeExtensions}
\end{figure}

The illustration shown in Figure~\ref{fig:ExtraeExtensions} depicts
where the instrumentation and sampling combined monitoring capabilities occurs during the application execution.
In the Figure, black markers represent instrumentation-based
points that record when a routine has started or finished executing
while red markers represent when the PMU has interrupted
the application for a PEBS sample after $X$ loads. The monitors
capture the value of the performance
counters and the top executing routine while
PEBS samples capture performance counters, a portion of the
call-stack and the PEBS record associated with the sample.

\subsubsection{Multiplexing sampling events}

Finally, the integration between Extrae and \texttt{perf} has also included multiplexing capabilities on the PEBS sampling.
That is, Extrae not only automatically changes the performance counters being collected at runtime but also can multiplex over PEBS sampling events.
Our implementation has covered the case in which Extrae monitors load and store instructions in a single run.
This has two major benefits.
First, the application only needs to run once rather than multiple times.
Second, this approach removes problems around matching addresses on different address spaces as a result of the introduction of the address randomization security mechanisms.

\subsection{Extensions to Folding}

The additional performance information captured by Extrae allows the Folding tool
to enrich its outputs (the reports and the synthetic trace-file).
The reports generated by the Folding correlate the progression within the source-code,
the performance and the address space in a single plot.
As the reports are limited by display properties, all the meaningful data is included into a Paraver trace-file for a quantitative and more detailed analysis.

In the report, the address space is partitioned based on the existing data objects and labeled
accordingly with the variable names (or call-stack) if available 
which allows the analyst to identify the data structures.
The report also includes memory references which shows how the data objects (including the wrapped data objects through the API extensions on Extrae) are accessed.
On multi-threaded/process applications, one report is generated for each executing thread/process.
This approach allows the analyst to explore each thread/process independently and to
learn whether different threads are accessing to shared or private variables,
although this exploration has to be done manually at the moment.
The forthcoming Section~\ref{subsec:ApplicationEvaluation_Stream} provides a full-featured analysis example.

Despite the Folding tool can combine the performance metrics from different processes
(as long as they refer to the same code), this is no longer possible when combining memory-related
information. The inclusion of the Address Space Layout
Randomization (ASLR) security techniques\footnote{\url{https://lwn.net/Articles/546686}}
leads to unique address spaces on each process even for the same binary and makes difficult to
combine the address space information from multiple processes. As a result,
the Folding tool only uses information from one process when exposing memory-related
information.
The ASLR mechanism required a manual matching of the data-objects in the initial
prototype when analyzing reports generated for load and store instructions independently.
This task was tedious, especially when applications refer to a large number of data objects.
However, the usage of the multiplexing capabilities for the PEBS sampling mechanism in Extrae
allows the Folding technique to depict the load and store references in the process address space
with a single execution.

\section{Application evaluation}
\label{sec:ApplicationEvaluation}

\subsection{Platform and Methodology}

We have evaluated several applications on the Jureca system~\cite{JURECA}
to show the usability of the extensions described above when exploring the load
and store references.
Each node of the system contains two Intel~Xeon E5-2680v3 (codename Haswell) 12-core processors
with hyper-threading enabled, for a total of 48 threads per node.
The nominal and maximum ``turbo'' processor frequencies are 2.50~GHz and 3.30~GHz,
respectively. The processor has three levels of cache with a
line size of 64 bytes: level 1 are two independent 8-way 32~KByte
caches for instructions and data, level 2 consists of an
8-way unified 256~KByte cache, and level 3 is a 20-way shared
unified 30,720~KByte cache. The system runs Linux~3.10.0,
has the GNU~v4.8.5 and Intel\textsuperscript{\textregistered} C and Fortran compilers~v15.0
and uses Intel\textsuperscript{\textregistered} MPI library~v5.1.
We have manually delimited with instrumentation points
the main iteration loop body of the respective applications.

With respect to the Extrae configuration, we have only captured dynamically-allocated objects that are larger or equal than 32~KByte.
Applications have been sampled every 137K~load and every 8231K~store instructions and the package has been configured to multiplex them every 15~seconds.
We use prime numbers to minimize the correlation between the sampling and the application periods.
The store sampling period is higher than the load sampling period because the Load-Latency feature already subsamples load instructions through a randomization tagging mechanism.
On the selected machine, the monitors (collecting time-stamp, call-stack and performance counters) take less than 2$\mu$s to execute for a measured overhead below 5\% on the presented experiments.

\subsection{Stream}
\label{subsec:ApplicationEvaluation_Stream}

For exemplification purposes, we have monitored the serial version of the Stream benchmark\footnote{Downloaded from \url{https://www.cs.virginia.edu/stream/FTP/Code/stream.c} with SHA-1 \texttt{afe4e58ec9ba61eba0b8b65cb24789295f8a539e}.}~\cite{STREAM}.
The benchmark has been compiled using the GNU compiler suite with each array being of size $N{=}2{\times}10^{7}$ elements.
As Stream accesses statically allocated variables through ordered linear accesses, we have modified the code so that:
(i) the \texttt{b} array is no longer a static variable but allocated by \texttt{malloc},
(ii) the scale kernel loads data from pseudo-random indices from the \texttt{c} array, and,
(iii) we have delimited the main application loop using the Extrae API calls.
Due to modification (ii), \texttt{scale} executes additional instructions and exposes less locality of reference, thus we have reduced the loop trip count in this kernel to $N{/}8$ to compensate its longest duration.
A simplified version of the code (in which kernel routines have been inlined) looks like:

{\small
\begin{program}
  \FOR i:=1 \TO NTIMES \DO \rcomment{! main loop}\\
     Extrae\_function\_begin() \rcomment{! Delimit begin body loop} \\
     \FOR j:=1 \TO N   \DO c[j]:=a[j]; \OD \rcomment{! Copy} \\
     \FOR j:=1 \TO N/8 \DO b[j]:=s*c[random(j)]; \OD \rcomment{! Scale} \\
     \FOR j:=1 \TO N   \DO c[j]:=a[j]+b[j]; \OD \rcomment{! Add} \\
     \FOR j:=1 \TO N   \DO a[j]:=b[j]+s*c[j]; \OD \rcomment{! Triad} \\
     Extrae\_function\_end() \rcomment{! Delimit end body loop} \\
  \OD 
  \label{alg:Stream_code}
\end{program}
}

\subsubsection{Description of the folding report}

\begin{figure}
	\centering
	\includegraphics[width=0.75\textwidth]{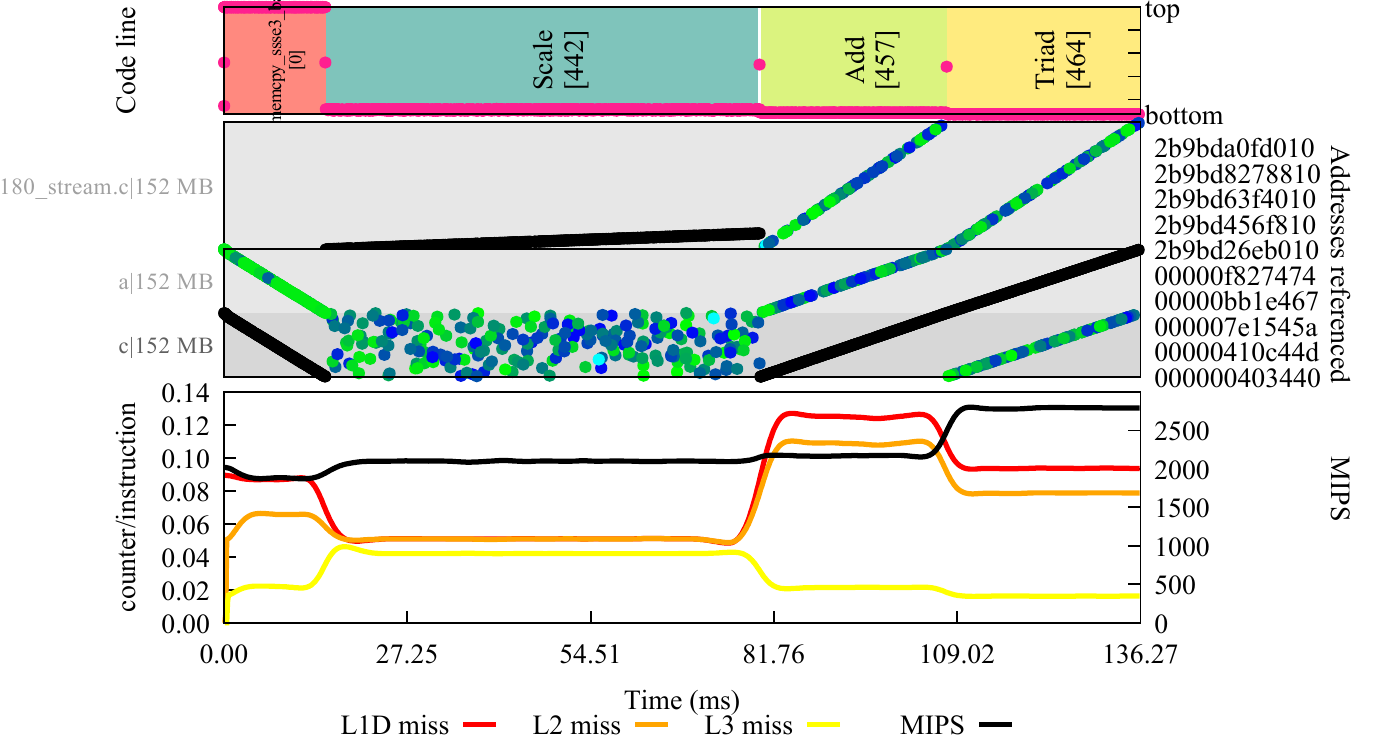}
	\caption{Analysis of the modified version of the Stream benchmark using the results from the Folding tool. There are triple correlation time-lines for the main iteration (from top to bottom): source code, addresses referenced and performance.}
	\label{fig:StreamExample}
\end{figure}

Figure~\ref{fig:StreamExample} shows the result of the extensions to the Folding mechanism.
The Figure consists of three plots: source code references
(top), address space load references (middle), and performance metrics (bottom).
In the source code profile each color represents the active routine
(identified by a label of the form X [$n$], where X refers to
the active routine, and $n$ refers the most observed code
line). Additionally, the purple dots represent a time-based
profile of the sampled code lines where the top (bottom) of
the plot represents the begin (end) of the source file. This
plot shows that the application progresses through four
routines (each representing a kernel) and that most of the
activity observed of each of these routines occurs in a tiny
amount of lines.
The second plot depicts the address space, including variable names of allocated
objects and memory references to the address space. On this plot,
the variables (either static or dynamically allocated) and their size are on the
left Y-axis, if any, and the right Y-axis shows the address space.
The dots in this plot show a time-based profile of the addresses referenced through load/store
instructions.
Load instructions are colored with a gradient that ranges from green to blue
referring to low and high access costs, respectively.
Store instructions are colored in black.
Finally, the third plot shows in black the achieved instruction (MIPS)
rate (referenced on the right Y-axis) within the instrumented
region, as well as the L1D, L2 and L3 cache misses per
instruction (on the left Y-axis) using red, orange and yellow,
respectively.
With this plot, the performance analyst can correlate different metrics of the performance
and see how they progress as the execution traverses code regions and accesses data objects.

\begin{table}
  \small
  \centering
  \caption{Classification and average costs of different accesses to the memory hierarchy per routine for the modified version of the Stream Benchmark.}
  \label{tab:ClassificationAndCosts_Stream}
    \begin{tabular}{l c c c c c c}
        \hlinethick
        \textbf{Routine} & \textbf{Metric} & \multicolumn{5}{c}{\textbf{Memory hierarchy part}} \\
                & & \textit{L1} & \textit{LFB} & \textit{L2} & \textit{L3} & \textit{DRAM} \\
        \hline
        \rowcolor{tabbg1}
        \multirow{2}{*}{Copy} & \% of load references   & 75.8\% & 22.5\% & 1.0\% & 0\% & 0.5\% \\
        \rowcolor{tabbg1}
        \multirow{-2}{*}{Copy} & Average cost (in cycles) & 7 & 28, 40 & 14 & \textit{n/a} & 400 \\

        \rowcolor{tabbg2}
        \multirow{2}{*}{Scale} & \% of load references   & 1.2\% & 80.5\% & 0\% & 4.6\% & 13.8\% \\
        \rowcolor{tabbg2}
        \multirow{-2}{*}{Scale} & Average cost (in cycles) & 7, 9 & 300, 340 & \textit{n/a} & 70 & 350, 800 \\

        \rowcolor{tabbg1}
        \multirow{2}{*}{Add} & \% of load references   & 3.8\% & 73.1\% & 0\% & 18.6\% & 3.8\% \\
        \rowcolor{tabbg1}
        \multirow{-2}{*}{Add}& Average cost (in cycles) & 7 & 50, 100 & \textit{n/a} & 70 & 400,440 \\

        \rowcolor{tabbg2}
        \multirow{2}{*}{Triad} & \% of load references  & 9.4\% & 74.2\% & 3.9\% & 8.9\% & 3.4\% \\
        \rowcolor{tabbg2}
        \multirow{-2}{*}{Triad}& Average cost (in cycles) & 7 & 74, 108 & 19 & 84 & 350 \\
        \hlinethick
    \end{tabular}
\end{table}

\subsubsection{Analysis of the folding report}

We outline several phenomena exposed from Figure~\ref{fig:StreamExample}.
First, as expected, the access
pattern in the Scale kernel to the variable \texttt{c} shows a randomized
access pattern with lots of high-latency (blue) references
while storing a portion of the memory allocated
in line 181 from file \texttt{stream.c} (the original variable \texttt{b}).
The straight lines formed by the references in the rest of the
routines denote that they linearly advance and thus expose
spatial locality, and also the greenish color indicates that
these references take less time to be served. Second, the instructions
within routines \texttt{Add} and \texttt{Triad} reference two addresses
per instruction on average, the loaded data comes
from two independent variables (or streams) simultaneously,
and their accesses go from low to high addresses honoring
the user code. Finally and surprisingly, the \texttt{Copy} routine accesses
the array in a downwards direction although the loop is written
with its index going upwards. This effect occurs because
the compiler has replaced the loop by a call to
\texttt{memcpy} (from glibc 2.17) that reverses the loop traversal
and uses SSSE3 vector instructions
(through the actual implementation \texttt{memcpy\_ssse3\_back}).
We observe that \texttt{Triad} and \texttt{Copy} benchmarks
achieve the highest and lowest MIPS rates, respectively. The
low MIPS rate in \texttt{Copy} may be explained because the execution of
vector instructions and these instructions take more
cycles to complete, but as a single instruction operates on multiple data,
it finalizes faster. Additionally, we expected a noticeable
difference regarding MIPS in \texttt{Scale} due to the introduction of
the random access to the variable. However, we observe that the instruction rate
is not significantly different between the kernels.
This happens because the \texttt{random()} function is inlined 
and avoids accessing memory by means of registers; thus, the additional instructions
do not miss in the cache and reduce the cache miss ratio per
instruction. Globally speaking, we notice that the L2 cache miss ratio
is similar to the L1D cache miss ratio.
This effect suggests that L2 provides little benefit
for this benchmark because L2 is not sufficiently large to keep the working
set. More specifically, we observe in the Scale kernel that the
L1D, L2 and L3 miss ratios are very similar (about 5\%) indicating
that each instruction that misses on L1D is likely to miss
on L2 and L3, as a result of the low temporal locality.
In addition, we can estimate the used memory bandwidth used in kernels that linearly access to variables (such as \texttt{Copy}, \texttt{Add} and \texttt{Triad}) if we consider that the whole variable is traversed (\textit{i.e.} the loop has 1-stride access).
Given these assumptions, the estimates indicate that \texttt{Copy} and \texttt{Triad} may use 20097 and 15263~MB/s of the memory bandwidth.
While these numbers are far from the nominal maximum memory bandwidth (68~GB/s\footnote{\url{http://ark.intel.com/products/81908/Intel-Xeon-Processor-E5-2680-v3-30M-Cache-2_50-GHz}}) for a single socket, the benchmark ran with one thread/process only and thus it is unlikely that it saturates the memory bus.

\subsubsection{Detailed analysis of the synthetic trace-file}

We have also explored the synthetic trace-files generated by
the Folding tool using Paraver. Table~\ref{tab:ClassificationAndCosts_Stream} summarizes
these results by showing the proportion of memory accesses
to the different parts of the memory hierarchy as well as the
average cost when accessing each part depending on the
active routine. Our first observation is the important
contribution of the Line-Fill Buffer (LFB) in terms of
percentage of accesses in terms of the average cost
in cycles.
The LFB is a buffer that keeps track of already requested cache-lines.
So memory references served by the LFB refer to load instructions that are
initiated by earlier and still incomplete instructions, thus exposing locality.
Consequently, the reported cost depends on the distance between load instructions and
the service time.
We highlight that LFB and DRAM costs show multi-modal
behaviors with high variability. For instance, in the Scale kernel,
data coming from DRAM takes either 350 or 800~cycles.
It is also worth mentioning that DRAM and LFB provide about 13.8\% and 80.5\% of the data to the \texttt{Scale} routine, respectively indicating a poor efficiency of the L1, L2 and L3 caches as a result of adding a random indirection.

\subsection{Lulesh v2.0}

\begin{table}
  \small
  \centering
	\rowcolors{2}{tabbg1}{tabbg2}
	\caption{Association for the labels shown in the Figure~\ref{fig:sub:Lulesh_Folding_Results_original}, including the most observed code line (MOCL) for each region.}
	\label{tab:Lulesh_Code_Association}
	    \begin{tabular}{l c c c c}
	    \hlinethick
	    \multicolumn{2}{l}{\textbf{Label}} & \textbf{User function} & \textbf{MOCL} & \textbf{Duration} \\
	    \hline
		\rowcolor{tabbg1}
	    \multirow{2}{*}{A}  & \textcolor{blue}{a1} & \texttt{CalcVolumeForceForElems} & 1105 & 268~ms \\
		\rowcolor{tabbg1}
	    \multirow{-2}{*}{A} & \textcolor{blue}{a2} & \texttt{CalcVolumeForceForElems} & 1121 & 268~ms \\
	    B & & \texttt{CalcHourglassControlForElems} & 1072 & 208~ms \\
	    C & & \texttt{LagrangeNodal} & 1263 & 80~ms \\
	    D & & \texttt{CalcLagrangeElements} & 1609 & 258~ms \\
	    E & & \texttt{CalcQForElems} & 1998 & 241~ms \\
	    F & & \texttt{ApplyMaterialPropertiesForElems} & 2424 & 616~ms \\
	    \hlinethick
	    \end{tabular}
\end{table}

\begin{table}
  \small
  \centering
  \caption{Classification and average costs of different accesses to the memory hierarchy per routine for the Lulesh benchmark.}
  \label{tab:ClassificationAndCosts_Lulesh}
    \begin{tabular}{l c c c c c c c c}
        \hlinethick
		\multicolumn{2}{l}{\textbf{Region} Subregion} & \textbf{Metric} & \multicolumn{5}{c}{\textbf{Memory hierarchy part}} \\
		        & & & \textit{L1} & \textit{LFB} & \textit{L2} & \textit{L3} & \textit{DRAM} \\
		\hline
		\rowcolor{tabbg1}
		                    & \multirow{2}{*}{\textcolor{blue}{a1}} & \% of load references & 99.66\% & 0.05\% & 0.28\% & 0\% & 0\% \\
		\rowcolor{tabbg1}
		\multirow{2}{*}{A}  & \multirow{-2}{*}{\textcolor{blue}{a1}} & Average cost (in cycles) & 7 & 25 & 14 & \textit{n/a} & \textit{n/a} \\

		\rowcolor{tabbg1}
		\multirow{-2}{*}{A} & \multirow{2}{*}{\textcolor{blue}{a2}} & \% of load references & 99.59\% & 0.14\% & 0.20\% & 0.06\% & 0\% \\
		\rowcolor{tabbg1}
		                    & \multirow{-2}{*}{\textcolor{blue}{a2}} & Average cost (in cycles) & 7 & 20 & 14 & 80 & \textit{n/a} \\

		\rowcolor{tabbg2}
		\multirow{2}{*}{B} & &  \% of load references & 98.37\% & 0.54\% & 1.00\% & 0.07\% & 0\% \\
		\rowcolor{tabbg2}
		\multirow{-2}{*}{B} & &  Average cost (in cycles) & 7 & 21 & 14 & 49 & \textit{n/a} \\

		\rowcolor{tabbg1}
		\multirow{2}{*}{C} & &  \% of load references & 98.80\% & 1.19\% & 0\% & 0\% & 0\% \\
		\rowcolor{tabbg1}
		\multirow{-2}{*}{C} & &  Average cost (in cycles) & 7 & 15 & \textit{n/a} & \textit{n/a} & \textit{n/a} \\

		\rowcolor{tabbg2}
		\multirow{2}{*}{D} & & \% of load references & 99.44\% & 0.12\% & 0.36\% & 0\% & 0.06\% \\
		\rowcolor{tabbg2}
		\multirow{-2}{*}{D} & & Average cost (in cycles) & 7 & 14 & 14 & \textit{n/a} & 350 \\

		\rowcolor{tabbg1}
		\multirow{2}{*}{E} & & \% of load references & 98.13\% & 0.7\% & 1.16\% & 0\% & 0\% \\
		\rowcolor{tabbg1}
		\multirow{-2}{*}{E} & & Average cost (in cycles) & 7 & 14 & 14 & \textit{n/a} & \textit{n/a} \\

		\rowcolor{tabbg2}
		\multirow{2}{*}{F} & & \% of load references & 96.55\% & 2.48\% & 0.45\% & 0.16\% & 0.33\% \\
		\rowcolor{tabbg2}
		\multirow{-2}{*}{F} & & Average cost (in cycles) & 7 & 95, 230 & 14, 19 & 109 & 300, 600 \\
        \hlinethick
    \end{tabular}
\end{table}

\begin{table}
  \small
  \centering
    \rowcolors{2}{tabbg2}{tabbg1}
	\centering
	\caption{Top 5 referenced variables in Lulesh identified by their allocation call-site.}
    \label{tab:Top_referenced_variables_Lulesh}
    \begin{tabular}{l c c r}
    \hlinethick
    \textbf{Allocation site} & \textbf{Size} & \textbf{\% of references} & Comment \\
    \hline
    lulesh.h 156 & 7~MB  & 1.23\% & coordinates \\
    lulesh.h 163 & 27~MB & 1.20\% & node list \\
    lulesh.h 143 & 7~MB  & 0.91\% & forces \\
    lulesh.h 154 & 7~MB  & 0.89\% & accelerations\\
    lulesh.h 147 & 7~MB  & 0.74\% & velocities \\
    \hlinethick
    \end{tabular}
\end{table}

\begin{figure}
	\centering
	\subfloat[Folding results for the main computation region of the Lulesh 2.0 benchmark.]
	{
		\includegraphics[width=\columnwidth]{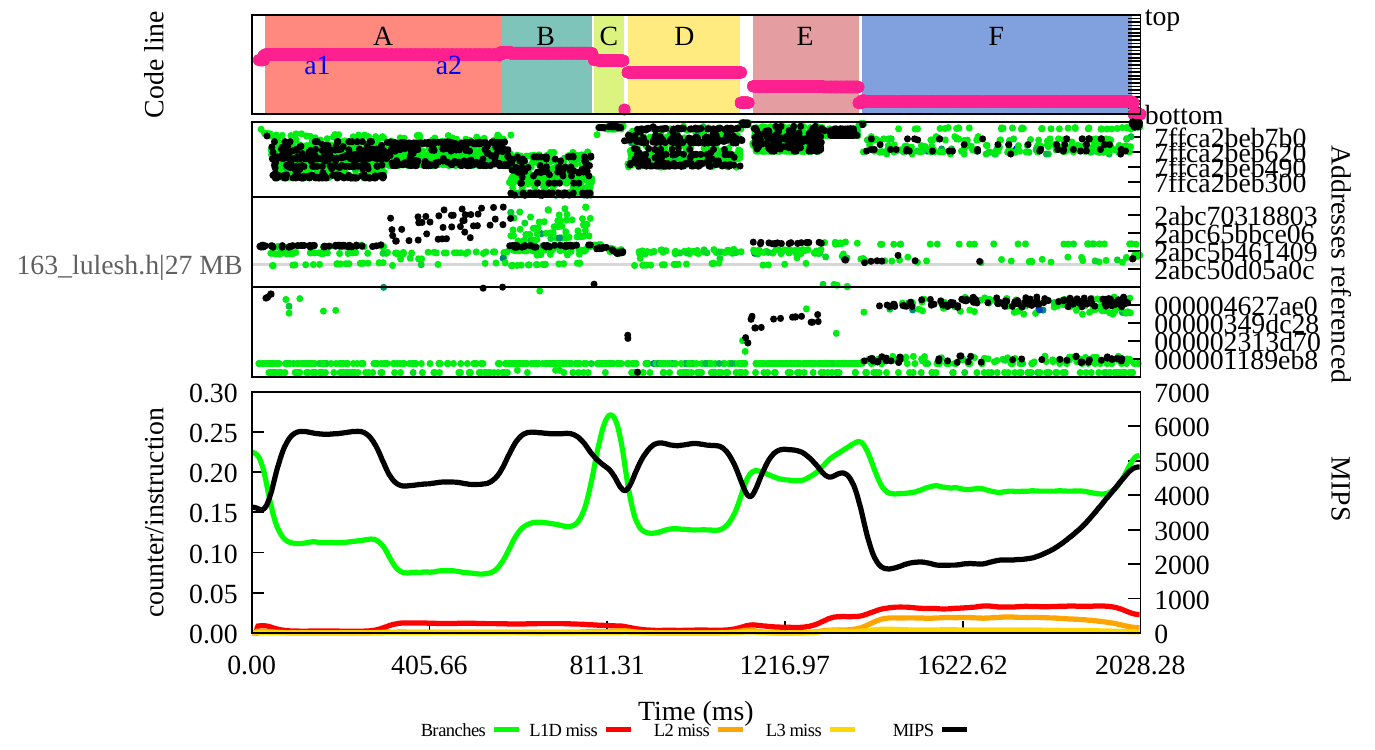}
		\label{fig:sub:Lulesh_Folding_Results_original}
	}
	\\
	\subfloat[Folding results for the main computation region of the Lulesh 2.0 benchmark after the code modifications (at the same time-scale as in Figure~\ref{fig:sub:Lulesh_Folding_Results_original}).]
	{
		\includegraphics[width=\columnwidth]{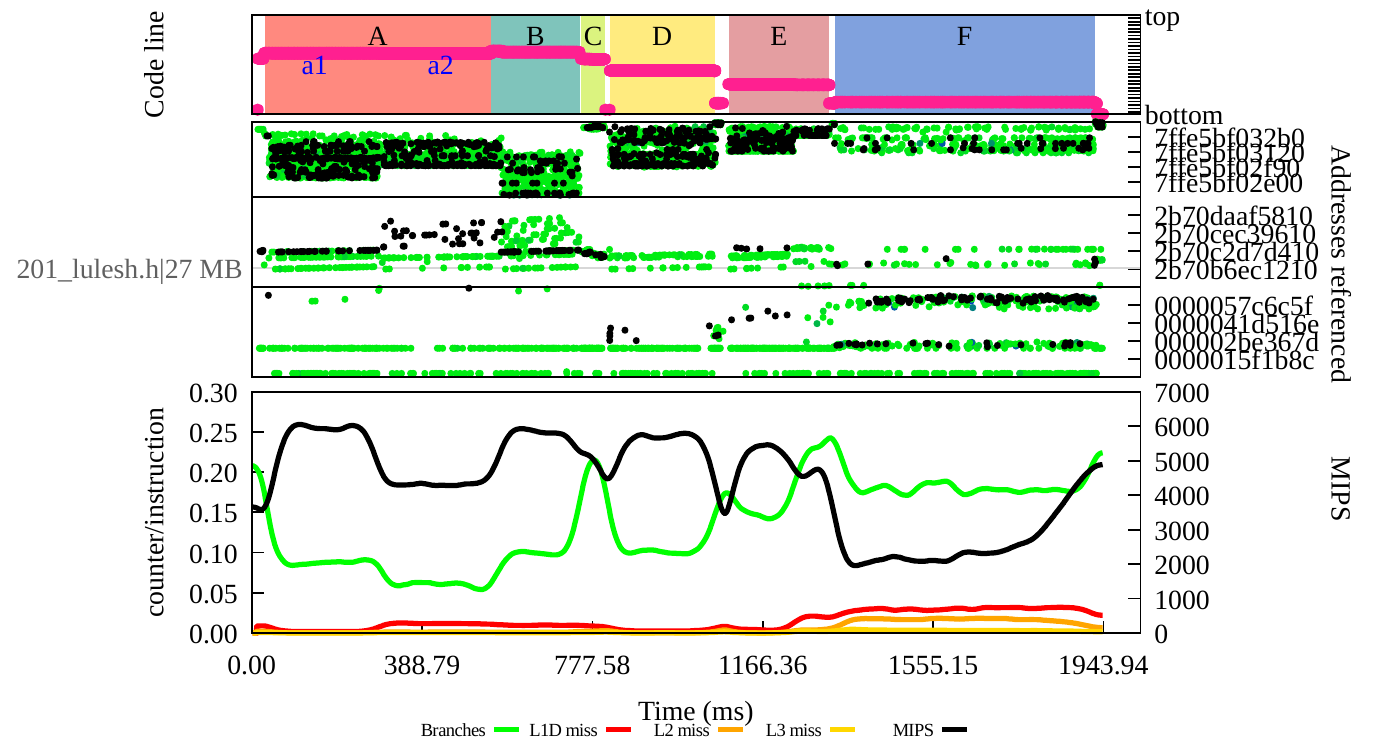}
		\label{fig:sub:Lulesh_Folding_Results_optimized}
	}
	\caption{Analysis of Lulesh 2.0.}
	\label{fig:Lulesh_Analysis}
\end{figure}

The Livermore Unstructured Lagrange Explicit Shock Hydrodynamics
(LULESH) proxy application~\cite{LULESH} is a representative
of simplified 3D Lagrangian hydrodynamics on an unstructured
mesh. We have compiled the reference code of the application\footnote{Downloaded from \url{https://codesign.llnl.gov/lulesh/lulesh2.0.3.tgz} with SHA-1 \texttt{541763c5015d094c667a79b004c22a78164fa4a4}.}
using the Intel compiler suite with the \texttt{-O3 -xAVX -g}
compilation flags. We have delimited with instrumentation points
the main iteration loop of the application and
executed it using 27 MPI processes on two nodes of Jureca with
a problem size $96^{3}$ for 200 iterations.

\subsubsection{Analysis of the folding report}

Figure~\ref{fig:sub:Lulesh_Folding_Results_original} shows the evolution of the code regions, accesses to
the address space and the performance within the main iteration
of the benchmark.
Due to lack of space in the plot, we have added labels (A-F) manually and the correspondence between the labels and Table~\ref{tab:Lulesh_Code_Association} shows the name of the routines.
The main loop traverses seven application regions (A-F), in which A is divided into two phases (\texttt{a1} and \texttt{a2}).
Regions A-E show good MIPS performance with IPC rates close to 2, while region F exposes a much lower performance.
Such a lower performance is correlated with an increase of the cache misses per instruction at all cache levels but still below 4\%.
The high part of the address
space refers to local variables allocated on the stack and the
rest of the allocations are performed through the \texttt{new} C++ language construct.
We observe a larger number of modifications within the stack region compared to the other parts of the address space.
We also notice that region \texttt{a2} writes on a region of the memory space (addresses prefixed by \texttt{0x2ab}) that is later read from region B and that phases E and F modify disjoint parts of the lower address space.
This information would be valuable when searching for parallelization opportunities using data-dependent task-based programming models.

\subsubsection{Detailed analysis of the synthetic trace-file}

Table~\ref{tab:ClassificationAndCosts_Lulesh} provides
detailed access statistics for the identified regions within
the main iteration. We see a general trend: L1 serves
most of the memory references, except for Region F
that shows a high value in the number of accesses provided by the
LFB and the access cost exposes a bi-modal behavior between
95 and 230~core cycles.

Further analysis with Paraver shows that there are many referenced memory objects and we tabulate the most referenced in Table~\ref{tab:Top_referenced_variables_Lulesh}.
The most referenced objects involve the nodelist (allocated in \texttt{lulesh.h} line 163), and the coordinates, the forces, the accelerations and the velocities of each element.
The four latter objects implement 3D floating-point arrays using 3 C++ vector containers (one per dimension) in a C++ class, such as a struct of arrays (SoA).
This storing method may not be efficient because the code pointed by Table~\ref{tab:Lulesh_Code_Association} shows concurrent accesses to the 3 dimensions per element, which may result in poor locality because memory references point to different containers.

We have changed the implementation of these 3D floating-point arrays  to an array of structs (AoS) aiming to increase the locality.
After applying the change, the Figure Of Merit (FOM) increased from $11891.71 z{/}s$ to $12414.23 z{/}s$, a 4.40\% increase.\footnote{As a side note, the benchmark includes a header file (\texttt{lulesh\_tuple.h}) to apply this change to additional structures than those we indicated but its usage reduced the FOM to 11081.57 $z/s$.}
Additionally, if we focus on the longest region (F) and explore the pointed code, we observe that it refers to an inline function invocation to the routine \texttt{EvalEOSForElems}.
The main loop of this routine consists of 3 inner loops that iterate over the number of elements and additional conditionals that may also execute an additional loop over all the elements.
By joining these loops to increase the locality and reduce the number of branch instructions, the FOM increased to $12480.03 z{/}s$ (a 4.80\% increase from baseline).
Figure~\ref{fig:sub:Lulesh_Folding_Results_optimized} shows the results for the Folding process when applied to the modified binary.
While the application behavior does not change abruptly, the overall MIPS rate is higher by 2\% responding to a L1 data-cache miss reduction by 9.6\%.
The optimized version also executes 2\% less number of instructions due to the reduction of the branch instructions executed (15.8\%).

\subsection{HPCG v3.0}

The High Performance Conjugate Gradient (HPCG) code benchmarks
computer systems based on a simple additive Schwarz,
symmetric Gauss-Seidel preconditioned conjugate gradient
solver~\cite{HPCG}. We have compiled the reference code of the application\footnote{Downloaded from \url{http://www.hpcg-benchmark.org/downloads/hpcg-3.0.tar.gz} with SHA-1 \texttt{39e1b7e45e67845f8551ff3c6ace5d3bc021524a}.}
using the Intel compiler suite with the \texttt{-O3 -xAVX -g} compilation flags. We have
executed the benchmark using 24 MPI processes on a single
node of the Jureca system using a problem size $nx{=}ny{=}nz{=}104$.
The application undergoes first a setup phase to test the system
resilience and ability to remain operational and then the
application runs the execution phase.
Here, we have ignored the setup phase and have delimited with instrumentation points the main execution phase only.

On a preliminary analysis of the application, we observed that most of the PEBS references were not associated to a memory object.
This occurred because the application allocates its data using many consecutive tiny allocations (10s to 100s of bytes) which are below the specified threshold (32~KByte) and thus not traced.
The data objects are allocated using two different mechanisms in lines 108-110 and 143 within the file \texttt{GenerateProblem\_ref.cpp}, respectively.
The first set of objects are allocated through the \texttt{new} C++ language construct while the second set are allocated through the \texttt{[]}-operator of the  C++ STL-based map structures.
We avoided creating huge event trace-files by grouping these allocations in two groups by using the new API call from Extrae.
Each grouped allocation covers the first and last addresses of all the included allocations.
Even though memory allocators may use different arenas (each on a different part of the address space) to reduce memory fragmentation, this approach served our purposes because the allocated regions were located in consecutive addresses.

\begin{table}
  \small
  \centering
  \caption{Code association for the labels shown in Figure~\ref{fig:sub:HPCG_Folding_Results_original} including the most observed code line (MOCL) for each region.}
  \label{tab:HPCG_code_Association}
    \rowcolors{2}{tabbg1}{tabbg2}
    \begin{tabular}{l c c c c}
    \hlinethick
    \multicolumn{2}{l}{\textbf{Label}} & \textbf{User function} & \textbf{MOCL} & \textbf{Duration}\\
    \hline
	\rowcolor{tabbg1}
    \multirow{2}{*}{A}  & \textcolor{blue}{a1} & \texttt{ComputeSYMGS\_ref} & 76 & 147~ms\\
	\rowcolor{tabbg1}
    \multirow{-2}{*}{A} & \textcolor{blue}{a2} & \texttt{ComputeSYMGS\_ref} & 95 & 143~ms\\

    B & & \texttt{ComputeSPMV\_ref} & 68 & 116~ms\\

    C & & \texttt{ComputeMG\_ref} & 47 & 96~ms\\

	\rowcolor{tabbg2}
    \multirow{2}{*}{D}  & \textcolor{blue}{d1} & \texttt{ComputeSYMGS\_ref} & 76 & 147~ms\\
	\rowcolor{tabbg2}
    \multirow{-2}{*}{D} & \textcolor{blue}{d2} & \texttt{ComputeSYMGS\_ref} & 95 & 143~ms\\

	\rowcolor{tabbg1}
    E & & \texttt{ComputeSPMV\_ref} & 68 & 136~ms \\
    \hlinethick
	\end{tabular}
\end{table}

\subsubsection{Analysis of the folding report}

Figure~\ref{fig:sub:HPCG_Folding_Results_original} shows the result of the folding tool when applied to the modified version of the HPCG benchmark and Table~\ref{tab:HPCG_code_Association} associates the code regions (A-E) shown in the Figure with the actual code.
We notice that each iteration consists of
two rounds of calls to \texttt{ComputeSYMGS\_ref} (labels A and D)
and \texttt{ComputeSPMV\_ref} (labels B and E) and in between there
is a call to \texttt{ComputeMG\_ref} (label C).
We identify linear accesses in the higher and lower part of the address
space. More precisely, regions A and D present a phase (labeled
as \texttt{a1} and \texttt{d1} in blue) that accesses the address space from
lower to upper addresses followed by a phase (labeled as \texttt{a2}
and \texttt{d2} also in blue) that accesses the address space from upper
to lower addresses. The lower to upper accesses represent one
forward sweep, while the upper to lower accesses represent
a backward sweep. It is worth to note that there are no stores
(\textit{i.e.} black points) in the lower part of the address space in the
execution phase, suggesting that data has been written in
an earlier application phase.

\begin{table}
  \small
  \centering
  \caption{Classification and average costs of different accesses to the memory hierarchy per routine for the HPCG Benchmark.}
  \label{tab:ClassificationAndCosts_HPCG}
    \begin{tabular}{l c c c c c c c}
        \hlinethick
		\multicolumn{2}{l}{\textbf{Region} Subregion} & \textbf{Metric} & \multicolumn{5}{c}{\textbf{Memory hierarchy part}} \\
		        & & & \textit{L1} & \textit{LFB} & \textit{L2} & \textit{L3} & \textit{DRAM} \\
		\hline
		\rowcolor{tabbg1}
		                   & \multirow{2}{*}{\textcolor{blue}{a1}} & \% of load references & 58.8\% & 30.7\% & 1.8\% & 1.6\% & 1.2\% \\
		\rowcolor{tabbg1}
		\multirow{2}{*}{A} & \multirow{-2}{*}{\textcolor{blue}{a1}} & Average cost (in cycles) & 7 & 15, 70 & 14 & 50 & 350, 450 \\

		\rowcolor{tabbg1}
		\multirow{-2}{*}{A} & \multirow{2}{*}{\textcolor{blue}{a2}} & \% of load references & 61.8\% & 21.8\% & 2.2\% & 1.6\% & 2.0\% \\
		\rowcolor{tabbg1}
		                    & \multirow{-2}{*}{\textcolor{blue}{a2}} & Average cost (in cycles) & 7 & 15 & 14 & 65 & 350, 700 \\

		\rowcolor{tabbg2}
		\multirow{2}{*}{B} & &  \% of load references & 58.9\% & 30.9\% & 1.5\% & 1.6\% & 1.2\% \\
		\rowcolor{tabbg2}
		\multirow{-2}{*}{B} & & Average cost (in cycles) & 7 & 60 & 14 & 65 & 540 \\

		\rowcolor{tabbg1}
		\multirow{2}{*}{C} & & \% of load references & 62.2\% & 24.8\% & 2.0\% & 1.4\% & 2.0\% \\
		\rowcolor{tabbg1}
		\multirow{-2}{*}{C} & & Average cost (in cycles) & 7 & 70 & 14 & 110 & 300, 450 \\

		\rowcolor{tabbg2}
		                    & \multirow{2}{*}{\textcolor{blue}{d1}} & \% of load references & 59.9\% & 30.5\% & 2.2\% & 1.0\% & 1.4\% \\
		\rowcolor{tabbg2}
		\multirow{2}{*}{D}  & \multirow{-2}{*}{\textcolor{blue}{d1}} & Average cost (in cycles) & 7 & 15, 70 & 14 & 50 & 350 \\

		\rowcolor{tabbg2}
		\multirow{-2}{*}{D} & \multirow{2}{*}{\textcolor{blue}{d2}} & \% of load references & 62.7\% & 22.5\% & 1.7\% & 1.0\% & 2.3\% \\
		\rowcolor{tabbg2}
		                    & \multirow{-2}{*}{\textcolor{blue}{d2}} & Average cost (in cycles) & 7 & 15 & 15 & 50 & 350 \\

		\rowcolor{tabbg1}
		\multirow{2}{*}{E} & & \% of load references & 57.1\% & 33.8\% & 1.7\% & 1.5\% & 0.5\% \\
		\rowcolor{tabbg1}
		\multirow{-2}{*}{E} & & Average cost (in cycles) & 7 & 15, 50 & 16 & 70 & 270, 330 \\
        \hlinethick
    \end{tabular}
\end{table}

From the performance perspective, the code does not exceed 1500~MIPS (IPC of 0.6 at the nominal frequency).
The transitions between phases expose higher instruction and branch instruction rates and a cache miss reduction.
Within routines, the instruction rate increases marginally when the application moves from forward sweep to backward sweep (regions \texttt{a1} to \texttt{a2} and \texttt{d1} to \texttt{d2}).

The report indicates that \texttt{a1} and \texttt{a2} traverse the whole data structure, the approximations for the memory bandwidth while traversing the structure are 4197~MB/s and 4315~MB/s, respectively.
In comparison, the observed bandwidth while traversing the same structure in region \texttt{B} achieves 6427~MB/s.
These values are smaller than the bandwidth observed in the Stream example but we have to consider that (i) there were 24 MPI ranks running in the same node and competing for the bandwidth and (ii) the report provides the performance for a single process/thread.

\subsubsection{Detailed analysis of the synthetic trace-file}

Table~\ref{tab:ClassificationAndCosts_HPCG} shows that the backward sweeps hit approximately 3\% more
in L1D compared to forward sweeps, about 8-9\% less in LFB
and an additional 1\% of the references miss in all the caches
and have to go to DRAM.
The data provided by LFB presents multi-modal cost access that is difficult to characterize.

\begin{figure}
	\centering
 	\subfloat[Folding results for the main computation region of the HPCG 3.0 benchmark.]
 	{
 		\includegraphics[width=\columnwidth]{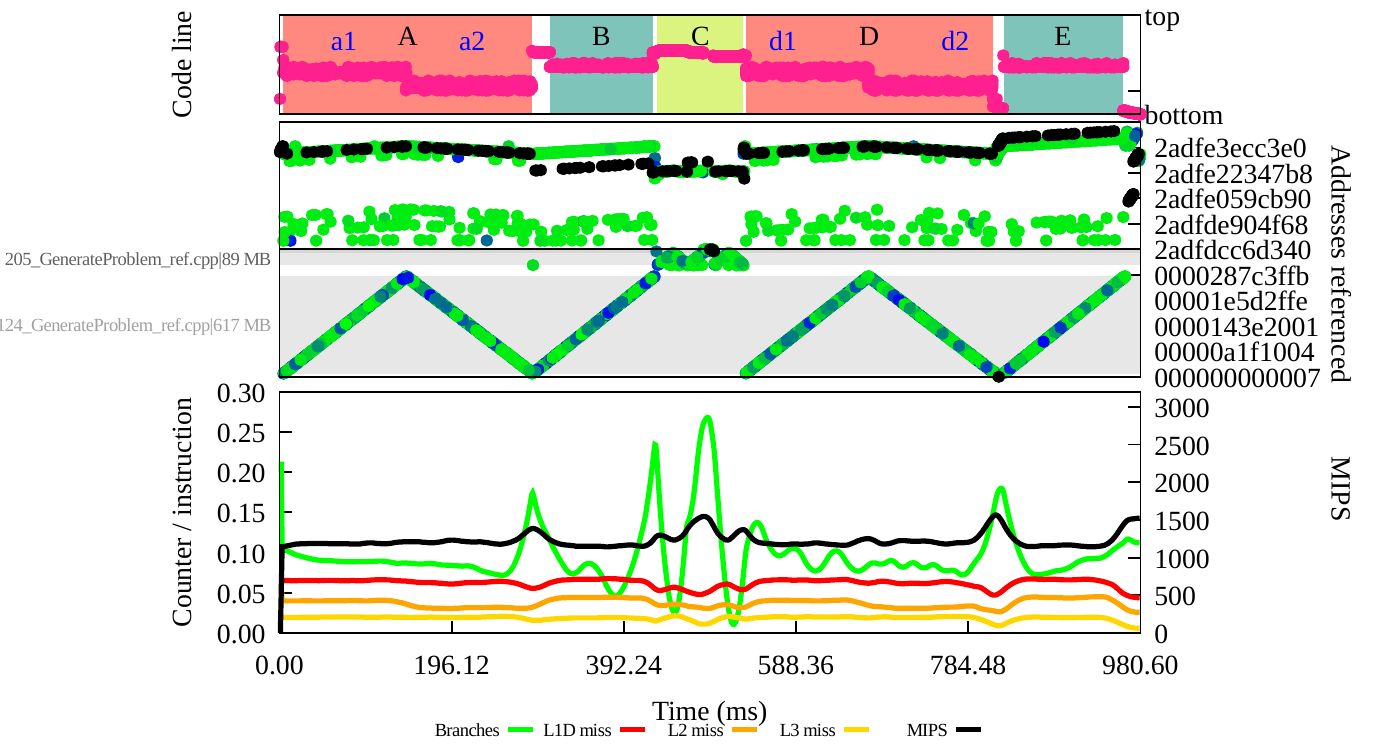}
 		\label{fig:sub:HPCG_Folding_Results_original}
 	}
 	\\
 	\subfloat[Folding results for the main computation region of the HPCG 3.0 benchmark after the code modifications (at the same time-scale as in Figure~\ref{fig:sub:HPCG_Folding_Results_original}).]
 	{
 		\includegraphics[width=\columnwidth]{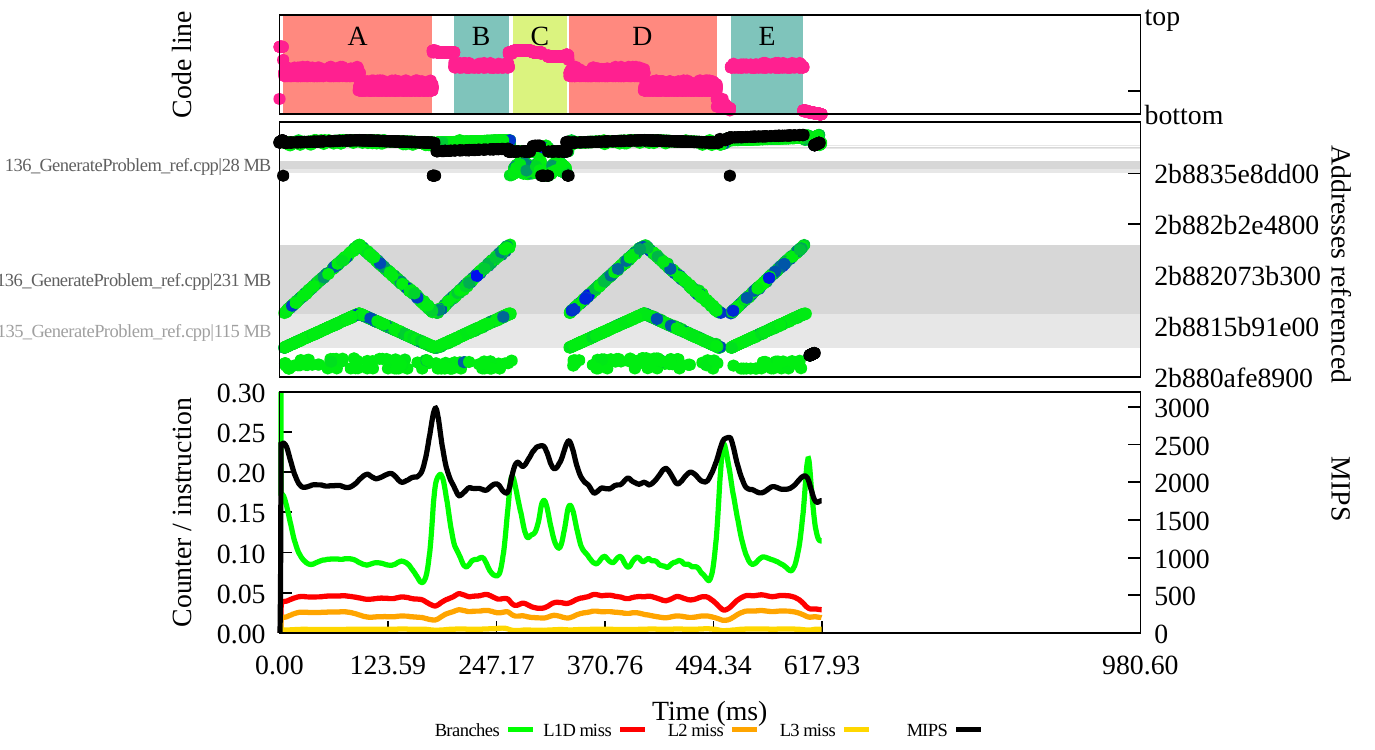}
 		\label{fig:sub:HPCG_Folding_Results_optimized}
 	}
 	\caption{Analysis of HPCG 3.0.}
 	\label{fig:HPCG_Analysis}
\end{figure}

\begin{table}
    \small
	\centering
	\caption{Top 5 referenced variables in HPCG identified by their allocation call-site.}
    \label{tab:Top_referenced_variables_HPCG}
	    \begin{tabular}{l c c@{} r}
	    \hlinethick
	    \textbf{Allocation site} & \textbf{Size} & \textbf{\% of refs} & Comment \\
	    \hline
			\rowcolor{tabbg1}
	    GenerateProblem\_ref.cpp 124\footnotemark[12] & 617~MB &46.21\% & sparse matrix \\
			\rowcolor{tabbg2}
	    \multirow{2}{*}{GenerateProblem\_ref.cpp 205\footnotemark[13]} & \multirow{2}{*}{89~MB} & \multirow{2}{*}{4.96\%} & global/local maps \\
			\rowcolor{tabbg2}
	    \multirow{-2}{*}{GenerateProblem\_ref.cpp 205\footnotemark[13]} & \multirow{-2}{*}{89~MB} & \multirow{-2}{*}{4.96\%} & local/global maps \\
			\rowcolor{tabbg1}
	    GenerateProblem\_ref.cpp 124\footnotemark[12] & 78~MB & 4.74\% & sparse matrix \\
			\rowcolor{tabbg2}
	    GenerateProblem\_ref.cpp 124\footnotemark[12] & 10~MB & 0.70\% & sparse matrix \\
			\rowcolor{tabbg1}
	    GenerateProblem\_ref.cpp 124\footnotemark[12] & 1517~kB & 0.44\% & sparse matrix \\
	    \hlinethick
	    \end{tabular}
\end{table}

\footnotetext[12]{This line corresponds to lines 108-110 in the original code.}
\stepcounter{footnote}
\footnotetext[13]{This line corresponds to lines 132-134 in the original code.}
\stepcounter{footnote}

The results shown in Table~\ref{tab:Top_referenced_variables_HPCG} prove the high number of references to the memory objects that we wrapped.
It is known that the C library does not provide consecutive addresses to consecutive allocation calls because of (i) internal book-keeping to track free blocks, (ii) minimum allocation size, and (iii) alignment padding if needed.
Consequently, the object allocated in a single allocation will be more compact than the object allocated through small allocation calls and thus it is likely to expose better spatial locality.
With this in mind, we changed the allocation of the data objects to minimize the number of allocations and the results.
Using this modified version of the code, the FOM reported by the benchmark increased from 9.95 to 15.64 GFLOP/s (57\% higher than the original) and the performance results of the new version are shown in Figure~\ref{fig:sub:HPCG_Folding_Results_optimized}.
The Figure shows that the main computation phase on the new version lasts approximately 618~ms (37\% less) and that cache misses have decreased (for instance, L1D misses [in red] are always below 5\%).
Regarding the address space, we observe the following.
First, the (wrapped) memory object allocated in \texttt{GenerateProblem\_ref.cpp} line 124 split into two memory objects.
Second, the (wrapped) object allocated in the original version occupied 617~MB while the two objects on the newer version occupy 346~MB (56\% of original size) demonstrating that the object is more packed and might expose better spatial locality.
Third, linear accesses that we recognized in the (wrapped) object are still visible in the two objects but there are concurrent linear accesses to both.
With respect to performance, we notice a higher MIPS rate (70\% increase compared to the original) due to improved cache usage that largely compensates the additional instructions executed (7\%).
Regions \texttt{a1}, \texttt{a2} and \texttt{B} show less bandwidth usage (3844, 4325 and 5580~MB/s respectively) than the previous version which means that there is room for growth.

\section{Related work}
\label{sec:RelatedWork}

This section describes earlier approaches related to performance analysis tools that have focused to some extent on the analysis of data structures and the efficiency achieved while accessing to them.
We divide this research into two groups depending on the mechanism used to capture the addresses referenced by the load/store instructions.

The first group includes tools that instrument the application instructions to obtain the referenced addresses.
MemSpy~\cite{MemSpy} is a prototype tool to profile applications on a system simulator that introduces the notion of data-oriented, in addition to code oriented, performance tuning.
This tool instruments every memory reference from an application run and leverages the references to a memory simulator that calculates statistics such as cache hits, cache misses, \textit{etc.} according to a given cache organization.
SLO~\cite{SLO} suggests locality optimizations by analyzing the application reuse paths to find the root causes of poor data locality.
This tool extends the GCC compiler to capture the application's memory accesses, function calls, and loops in order to track data reuses, and then it analyzes the reused paths to suggest code loop transformations.
MACPO~\cite{MACPO} captures memory traces and computes metrics for the memory access behavior of source-level data structures.
The tool uses PerfExpert~\cite{Burtscher_2010} to identify code regions with memory-related inefficiencies, then employs the LLVM compiler to instrument the memory references, and, finally, it calculates several reuse factors and the number of data streams in a loop nest.
Intel\textsuperscript{\textregistered} Advisor is a component from the Intel\textsuperscript{\textregistered} Parallel Studio XE~\cite{IntelParallelStudio} that provides users insights on applications' vectorization.
It relies on PIN~\cite{PIN} to instrument binaries and precisely correlates memory access on user selected routines with source-code.
Tareador~\cite{Tareador} is a tool that estimates how much parallelism can be achieved in a task-based data-flow programming model.
The tool employs dynamic instrumentation to monitor the memory accesses of delimited regions of code in order to determine whether they can simultaneously run without data race conditions, and then it simulates the application execution based on this outcome.
EVOP is an emulator-based data-oriented profiling tool to analyze actual program executions in a system equipped only with a DRAM-based memory~\cite{EVOP}.
EVOP uses dynamic instrumentation to monitor the memory references in order to detect which memory structures are the most referenced and then estimate the CPU stall cycles incurred by the different memory objects to decide their optimal object placement in a heterogeneous memory system by means of the \textsl{dmem\_advisor} tool~\cite{DMEM_ADVISOR}.
ADAMANT~\cite{ADAMANT} uses the PEBIL instrumentation package~\cite{PEBIL} and includes tools to characterize application data objects, to provide reports helping on algorithm design and tuning by devising optimal data placement, and to manage data movement improving locality.

The second group of tools take benefit of hardware mechanisms to sample addresses referenced when processor counter overflows occur and estimate the accesses weight from the sample count.
The Oracle Developer Studio~\cite{ORACLE_DEVELOPER_STUDIO} (formerly known as Sun ONE Studio) incorporates a tool to explore memory system behavior in the context of the application's data space~\cite{Memory_Profiling_using_Hardware_Counters}.
This extension brings the analyst independent and uncorrelated views that rank program counters and data objects according to hardware counter metrics and it shows metrics for each element in data object structures.
HPCToolkit has been recently extended to support data-centric profiling of parallel programs~\cite{HPCTOOLKIT_MEM}, providing a graphical user interface that presents data- and code-centric metrics in a single panel, easing the correlation between the two.
Roy and Liu developed StructSlim~\cite{STRUCTSLIM} on top of HPCToolkit to determine memory access patterns to guide structure splitting.
Gim\'enez \textit{et al.} use PEBS to monitor load instructions that access addresses within memory regions delimited by user-specified data objects and focusing on those that surpass a given latency~\cite{GimenezGRBJBH14}.
Then, they associate the memory behavior with semantic attributes, including the application context which is shown through the MemAxes visualization tool.

The BSC tools for the memory exploration adopt a hybrid approach combining PEBS-based sampling and minimal instrumentation usage and its main difference from existing tools relies on the ability to report time-based memory access patterns, in addition to source code profiles and performance bottlenecks.
Regarding the monitoring mechanism, the tool brings two benefits.
First, limiting the instrumentation usage reduces the overhead suffered by the application and thus increases the representability of the performance results.
Second, the folding mechanism allows the analyst to blindly choose a sampling frequency because the mechanism gathers samples from repetitive code regions into a synthetic one, and consequently minimizes the number of application executions.
Regarding the results provided, the inclusion of the temporal analysis permits time-based studies such as detection of simultaneous memory streams, ordering accesses to the memory hierarchy, and even, insights for extracting parallelism through task-based data-flow programming models.
The results also allow manually estimating the memory bus bandwidth usage per variable on a give region of code on linear accesses.

\section{Conclusions}
\label{sec:Conclusions}

Memory hierarchies are getting complex and it is necessary to better understand the application behavior in terms of memory accesses to improve the application performance and prepare for future memory technologies.
The PEBS hardware infrastructure assists with sampling memory-related instructions and gathers valuable details about the application behavior.
We have described the latest extensions in the Extrae instrumentation package order to enable performance analysts to understand the application and system behavior in terms of memory accesses even for in-production optimized binaries.
The additional extension to the folding mechanism depicts the temporal evolution of the memory accesses in a compute region by using a coarse-grain non-intrusive sampling frequency and minimal instrumentation.
The usage of these tools results in thorough memory access patterns exploration on two state-of-the-art benchmarks without having to use high-frequency sampling and thus not incurring on large overheads.
The exploration included scan of the memory access patterns from a time perspective and the identification of the most dominant data streams and their temporal evolution along computing regions.
As a result of this exploration, we have proposed small changes to both of them that improved their performance.

In addition to the optimization efforts, application developers can use the presented tools to explore how the address space is being accessed and confirm if the results match their expectations.
For instance, the results for the modified Stream show that a user can identify the modification applied to the benchmark as well as the compiler decision to replace the source code by a \texttt{memcpy} call
that accesses the address space in reverse order compared to what the developer would expect.
Concerning Lulesh, the results show potential independent load and store accesses to the same parts of the address space by different routines which may be a valuable insight for using data-dependent task-based programming models.
Finally, the HPCG results show that the main routine traverses the address space two times (in a forward direction followed by a backward direction) and that a part of the address space is not modified.
HPCG also shows different performance values for forward and backward sweeps not only in cache miss ratios but also in the cost of providing data from memory.

Hardware architects may also find valuable insight in the results obtained.
One possible suggestion according to the Stream results would be to not cache in L2 given parts of the address space for a period of time with the consequent energy savings.
Additionally, the results for HPCG indicate that a portion of the address space is only read during the execution phase and thus this region may benefit from memory technologies where loads are faster than stores.

\section*{Acknowledgments}

This work has been performed in the Intel-BSC Exascale Lab.
We would like to thank Forschungszentrum J\"ulich for the compute time on the Jureca system.
This project has received funding from the European Union's Horizon 2020 research and innovation program under Marie Sklodowska-Curie grant agreement No. 749516.

\bibliographystyle{elsarticle-num}
\bibliography{bibliography}

\end{document}